# Strong Band-Edge Light Emission from InGaAs RTDs: Evidence for the Universal Nature of Resonant- and Zener- Co-Tunneling

E. R. Brown,[1*] W-D. Zhang,[1§] T. A. Growden,[2] P. R. Berger,[2] R. Droopad[3]

[1]Dept. of Physics, Wright State Univ., Dayton, OH 45435, USA
[2]Dept. of Electrical and Computer Engineering, Ohio State Univ., Columbus, OH 43210, USA
[3]Ingram School of Engineering, Texas State Univ., San Marcos, TX 78666, USA

## Abstract

We report strong light emission from a room-temperature n-type unipolar-doped $In_{0.53}Ga_{0.47}As$/AlAs double-barrier resonant-tunneling diode (DBRTD) precisely at the $In_{0.53}Ga_{0.47}As$ band-edge near 1650 nm. The emission characteristics are very similar to what was observed recently in GaN/AlN DBRTDs, both of which suggest that the mechanism for emission is cross-gap electron-hole recombination via resonant- and Zener co-tunneling of electrons, the latter mechanism generating the required holes. Analysis shows that because of the relatively small bandgap, the Zener tunneling probability can be large in this $In_{0.53}Ga_{0.47}As$/AlAs DBRTD, and is a mechanism that may have been overlooked in the longstanding literature. The universal nature of the co-tunneling is best supported by the factor $(E_G)^2/F$ in the Kane tunneling probability, which is nearly the same at the peak voltage of the $In_{0.53}Ga_{0.47}As$ and GaN DBRTDs.

[*] elliott.brown@wright.edu
[§] wzzhang@fastmail.fm

In recent research on n-type unipolar GaN/AlN DBRTDs, bright near-UV electroluminescence (EL) [1] was discovered in addition to a reproducible negative differential resistance (NDR) at room temperature [2-6]. Through spectral measurements, the UV emission was found to be centered at the GaN band-gap wavelength around 365 nm, and through recent noise measurements that the transport displayed normal shot noise except for a suppression effect associated with the resonant tunneling. These results combined with detailed quantum transport computation [1] suggested that the near-UV emission was by cross-gap radiative recombination between electrons accumulated on the emitter side of the device, and holes created in the same region generated by Zener tunneling, which is enabled in vertical GaN heterostructures by the huge polarization fields at the GaN/AlN heterointerfaces, leading to significant localized band bending[1, 7-10]. In this letter, we report on the first observation of the same mechanism for emission in an $In_{0.53}Ga_{0.47}As$/AlAs double-barrier RTD (DBRTD) but at the $In_{0.53}Ga_{0.47}As$ band-gap wavelength around 1650 nm. This discovery strongly suggests that the resonant- and Zener- co-tunneling of electrons is a universal feature of unipolar DBRTDs and, remarkably, one that has never been reported in the vast literature of resonant-tunneling diodes over the past $40^+$ years.

The DBRTD device under test was grown by molecular beam epitaxy as an $In_{0.53}Ga_{0.47}As$/AlAs heterostructure on a semi-insulating InP substrate with a layer structure and doping profile as shown in Fig. 1(a). Its active region is comprised of two unintentionally doped (UID) AlAs barriers (thickness=2.4 nm) separated by an undoped $In_{0.53}Ga_{0.47}As$ quantum-well (width = 4.4 nm) layer, such that a quasibound level $E_1$ occurs in the quantum well at an energy of ≈0.40 eV above the $In_{0.53}Ga_{0.47}As$ conduction band edge under zero bias. This rather high confinement energy compared to typical DBRTDs means that a large bias is required to reach the condition of negative differential resistance (NDR), especially under forward bias (positive on top contact), because of the 100-nm low-doped spacer layer on the top side that depletes and supports a large voltage drop and electric field. A local peak in current occurs at the start of the NDR region at 1.80 V, and a valley at the end of the NDR region at 2.65 V with a peak-to-valley current ratio (PVCR) of 9.2, as displayed in the experimental I-V curve of Fig. 1(b). This excellent PVCR is characteristic of InGaAs vs GaAs-based DBRTDs going back to their first demonstrations [11, 12]. Another important device metric is the peak current density $J_P$ which is $3.5×10^4$ A/cm² in this 9×6 µm² active-area device, making it a good RTD for fast electrical switching [13] amongst other possible high-speed applications.

Based on previous characterization of GaN/AlN RTDs, the set-up shown in Fig. 2(a) was used, consisting of a precision I-V probe station, a near-IR light-emission detector, and a near-IR fiber spectrometer. The ambient temperature was ≈295 K. The detector was a large-area Ge photodiode with spectral response between 800 and 1800 nm, and having a peak responsivity of 0.85 A/W at a wavelength of 1550 nm. Its input was coupled to the DBRTD with a short light pipe, and its output was dc coupled to a solid-state electrometer having a current noise floor of ~1 pA. Shown in Fig. 1(b) (right vertical axis) is the photocurrent from the Ge diode as a function of RTD bias voltage (L-V curve). The photocurrent from the electrometer rises significantly above the noise floor at a bias voltage of ≈1.0 V, and increases monotonically with higher voltage through the NDR region up to the valley point. Then there is a precipitous drop at the valley voltage followed by a slow increase above that. That is, the change in photocurrent in the NDR region is anticorrelated to the change in electrical current. This behavior is similar to that observed for the near-UV photocurrent from one of the GaN/AlN DBRTDs [6], but for reasons that are not yet understood.

The fiber spectrometer is a room-temperature InGaAs-array-grating instrument[14] sensitive between 880 and 1750 nm and has a programmable spectral resolution, chosen for the present experiments to be 0.5 nm. Plotted in Fig. 2(b) are the spectral emission curves plotted vs. wavelength and parameterized by bias voltage at $V_B$ = 1.7, 2.1, 2.5, and 3.0 V. The middle two bias points are in the NDR region, and the first and last points are just below and above it, respectively. All four curves show a peak emission λ around 1580 nm, and a long-wavelength cut-off behavior around 1670 nm. Superimposed in Fig. 2(b) is the $In_{0.53}Ga_{0.47}As$ band-edge wavelength reference, λ = 1684 nm corresponding to the band-gap energy of 0.736 eV at 295 K, and calculated with the Varshni formula [15, 16]. The intersection of this reference line with all four spectral curves in their steeply rising edge suggests that the observed emission is occurring at or near the $In_{0.5}Ga_{0.47}As$ band edge. However, the strongest curves in the NDR region (i. e. $V_B$=2.1, 2.5V) are distinctly asymmetric with short-wavelength tails that extend to 1300 nm, or less. In addition, the weaker emission curve at bias outside the NDR region (i. e. $V_B$=1.7, 3.0 V) also display short-wavelength tails, but appear more symmetric. The light emission with a UV-VIS fiber spectrometer was examined next. No peak feature was observed in the wavelength range of 200-800 nm. This and the IR spectrum suggest little possibility of recombination between confined electrons and holes in their respective potential wells.

To understand the emission process better, Fig. 2(c) shows the brightest of the emission curves plotted against the ideal spontaneous emission expression for a bulk semiconductor EL [17]:

$$S(\nu) = A\nu^2 (h\nu - E_G)^{1/2} \exp[(E_G - h\nu)/k_B T] \qquad (1)$$

where $E_G$ is the $In_{0.53}Ga_{0.47}As$ band gap [0.736 eV at 295 K], A is a frequency-independent constant, and no external cavity effects are included. The agreement is satisfactory on the low frequency (long-wavelength) end, but clearly Eqn 1 decays much faster than the experiment on the short-wavelength end. The experimental short-wavelength emission, along with the effect of device self-heating, are discussed below.

A simple qualitative model that explains the experimental data is shown schematically in Fig. 3(a). The spectra of Fig. 2(b) and Fig.2(c) clearly indicate that the emission is most likely free-carrier cross-gap recombination occurring at or near the $In_{0.53}Ga_{0.47}As$ band edge, which requires free holes. Judging from the threshold in emission shown in Fig. 1(b) just below 1.0 V bias, the likely generation mechanism for holes is the interband (Zener) tunneling mechanism of Fig. 3(a). The band-bending in Fig. 3(a) is such that electrons can readily flow by resonant tunneling from the emitter to the collector through the quantum-well quasibound level ($E_1$). Furthermore, the bias is large enough that unoccupied conduction band states on the collector side line up energetically with occupied valence band states on the emitter side, so that interband tunneling can occur while conserving energy and crystal momentum. The lowest possible threshold bias for this process is approximately the $In_{0.53}Ga_{0.47}As$ bandgap of ≈0.75 eV, which is reasonably close to the experimental threshold. Note that this model is subtly different than that proposed for the GaN/AlN DBRTD in Ref. 1 where the interband tunneling can occur from the valence-band quasibound level in the quantum well to the collector side, followed by tunneling of the holes to the emitter side where radiative recombination occurs. This is because in GaN, the key factor in the Zener tunneling is the huge interfacial polarization field [7, 1], whereas with In$_{0.53}$Ga$_{0.47}$As the key factor is the narrow bandgap.

To support the hypothesis that Zener tunneling is significant in the present structure, we present a calculation of the valence-to-conduction band tunneling probability according to the classic Kane expression:

$$T = \frac{\pi^2}{9} exp\left(-\frac{\pi^2 E_G^2 \cdot m}{2hP \cdot F}\right) \qquad (2)$$

where $m$ is the electron mass in vacuum, $h$ is Planck's constant, $P$ is the momentum matrix element between the valence and conduction band cell-periodic wavefunctions, generally expressed as $E_P \equiv P^2/2m$, and $F$ is the electric field in units of eV/cm [18-19]. Strictly speaking, Eqn 2 applies to interband tunneling in a uniform electric field, which is a reasonable approximation in the present DBRTD [Fig. 3(a)]. So in Fig. 3(b) we plot Eqn 2 assuming $E_G$ = 0.736 eV, $E_P$ = 25.3 eV [20], and as a function of $F$ between $1.0 \times 10^5$ and $3.0 \times 10^5$ eV/cm (this value of $E_G$ is established by device and bandgap modeling described below). The tunneling probability increases more than 6-orders-of-magnitude over this range of bias field, and is essentially a decaying exponential dependence on the length of the band-gap barrier given roughly as $L_B \approx E_G/F$. Between the bias voltage where we first see significant light emission, $V_B \approx 0.75$ eV, and the peak voltage $V_B$ = 1.75 V, we observe $T$ increase ~50 times from $2 \times 10^{-7}$ to $1 \times 10^{-5}$. While these values may at first seem small compared to the transmission probabilities for resonant tunneling, which routinely fall in the range 0.1 to 1.0, the overall Zener tunneling current also depends on the "supply function" of electrons occupying the valence band on the emitter side, which is very large because of the large effective density-of-states and the high Fermi occupancy factor.

The qualitative model of Fig. 3(a) can also explain the much broader experimental emission peak than described by Eqn 1. The holes created by Zener tunneling occupy a normal 3D valence band, without confinement, whereas the electrons are subject to occupation of the quasi-2D region in the undoped spacer layer adjacent to the barriers on the emitter side. This so-called "pre-well" has been the subject of debate over the years in the context of DBRTD design and speed limitations. Here, it can have the effect of smearing out the recombination spectral signature. A hole, such as that shown in Fig. 3(a) that is created at (or diffuses to) the peak of the band profile on the emitter side, will emit photons very close to the band edge. However, a hole that is created in the "pre-well" region can only recombine with electrons that occupy the pre-well quasibound level or above [shown as $E_C$ in Fig. 3(a)], which is significantly elevated above the conduction band edge. Therefore, the emitted photon will have energy significantly above $E_G$, consistent with the experimental data in Fig. 2(b), (c). Whether or not this simple model can explain the anti-correlation in the NDR region between the electrical current and photocurrent shown in Fig. 2(b) remains to be seen, as this is a more complicated effect involving coupled quasibound states.

To further emphasize the universal nature of the co-tunneling and enhance the accuracy of the analysis, Fig. 4 compares the physical characteristics of the In$_{0.53}$Ga$_{0.47}$As /AlAs emitter structure studied here to a GaN/AlN structure studied previously [1]. The band-bending plots in Figs 4(a) and (b) were computed as self-consistent solutions to the coupled Poisson-Schrödinger equations at a bias voltage just below the respective NDR regions [21]. The high electric field in the barrier region of the InGaAs structure, combined with its relatively narrow band gap, makes Zener tunneling a significant transport mechanism. The much greater polarization-induced electric field in the GaN/AlN again makes Zener tunneling plausible in spite of the much larger GaN bandgap. The essential tunneling parameters of Eqn 2 for In$_{0.53}$Ga$_{0.47}$As and GaN are listed in Table I. Of utmost importance are the electric fields at the peak voltage, $F_P$ [from Figs. 4(a) and (b)], $2\times10^5$ V/cm and $5\times10^6$ V/cm for the In$_{0.53}$Ga$_{0.47}$As and GaN RTDs, respectively. This large difference makes the factor $(E_G)^2/F$ in Eqn 2 remarkably close at $F = F_P$: $(E_G)^2/F_P = 2.7\times10^{-8}$ and $2.3\times10^{-8}$ for the InGaAs and GaN, respectively. The only other material-dependent factor in Eqn 2 is P, which is only ≈25% different between the two materials, and is similarly comparable amongst all the common semiconductors independent of bandgap [20].

Also included in Table I is the bandgap at the operating temperature of each device. The bandgap is calculated using the Varshni expression $E_G(T) = E_G(T=0) - \alpha T^2/(T+\beta)$ with parameters given in Table I. The operating temperature is estimated by T = 295 K + ΔT with ΔT ≈ $P_0 \cdot R_{TH}$, where $P_0$ is the dc power dissipation, and $R_{TH}$ is the thermal resistance, also included in Table I. As both devices were mesas having 54 μm$^2$ active area, the only difference in $R_{TH}$ is the higher thermal conductivity of the GaN-based device compared to the InP-based device in the heat "spreading" region below the mesa. Heat transport to above the mesa through the contact metal is negligible in comparison.

The EL spectra for both structures are plotted in Figs. 4(c) and (d) vs wavenumber ($\sigma$ [cm$^{-1}$]) on an identical scale, with 4(c) being the same data as in 2(b) at 1.7 V bias. Also plotted in Figs. 4(c) and (d) are the ideal EL curves according to Eqn 1 assuming for the In$_{0.53}$Ga$_{0.47}$As device: T = 318 K and E$_G$ = 0.729 ($\sigma_G$ = 5.883×10$^3$ cm$^{-1}$); and for the GaN device, T = 355 K and E$_G$ = 3.410 ($\sigma_G$ = 27.48×10$^3$ cm$^{-1}$). For the InGaAs device the experimental EL curve peaks well above (in $\sigma$) the maximum of its ideal EL spectra, so emits the majority of its radiation above the band-edge $\sigma_G$, consistent with the "pre-well" quantization effect described above. However, for the GaN device the experimental EL curve peaks close to the ideal-spectrum maximum and has a much broader width, such that the emission above and below the band gap are roughly equal. We again attribute the blue-shifted radiation to the "pre-well" quantization effect, which is strong in GaN as well as In$_{0.53}$Ga$_{0.47}$As. The red-shifted radiation is not as straightforward. In our previous analysis, the red-shift was obviated by renormalization of the GaN bandgap – an effect which decreases the bandgap energy in proportion to the free carrier concentration [6]. However, the lack of red-shifted radiation in the InGaAs device of Fig. 4 (a), even in the presence of the high accumulated electron density in the emitter region, suggests that bandgap renormalization may not be significant. Another possibility for the red-shifted GaN radiation is shallow traps that occur at the GaN emitter layer or at the GaN/AlN interfaces. This is supported by the experimental fact that the total GaN emission spectrum is significantly broader [FWHM =1060 cm$^{-1}$ in Fig. 4(d)] than the InGaAs spectrum [FWHM = 896 cm$^{-1}$ in Fig. 4(c)]. However, more research is necessary to resolve this discrepancy.

In conclusion, unipolar-doped double-barrier RTDs have now been shown to manifest electron resonant and Zener (interband) co-tunneling in two different materials systems, including the historic In$_{0.53}$Ga$_{0.47}$As/AlAs system described here for the first time. This mechanism creates holes at the emitter side of the structure where electrons are heavily accumulated, leading to band-to-band light emission. This has been an overlooked feature of RTDs since their invention in 1973 [22].

This work was supported by the U.S. National Science Foundation (under Grant #1711733), Program Director Dr. Dimitris Pavlidis.

| | Table I | | |
|---|---|---|---|
| | **Parameter** | **In$_{0.53}$Ga$_{0.47}$As on InP** | **GaN on GaN** |
| Physical | Mesa Area [um$^2$] | 54 | 54 |
| Electrical | I$_P$ [mA] | 17.1 | 23.0 |
| | V$_P$ [V] | 1.7 | 6.2 |
| | P$_0$ [mW] | 29.1 | 144.9 |
| Thermal | R$_{TH}$ [K/W] | 798 | 418 |
| | $\Delta$T [K] | 23.2 | 60.5 |
| Varshni | E$_G$(T=0) | 0.803 | 3.51 |
| | $\alpha$ [meV/K] | 0.400 | 0.909 |
| | $\beta$ [K] | 226 | 830 |
| | E$_G$(295K) [eV] | 0.736 | 3.44 |
| Zener | E$_G$(T + $\Delta$T) [eV] | 0.729 | 3.41 |
| | F$_P$ [V/m] | 2.0x10$^7$ | 5.0x10$^8$ |
| | E$_P$ [eV] | 25.3 | 20.2 |
| | (E$_G$)$^2$/F$_P$ | 2.7x10$^{-8}$ | 2.3x10$^{-8}$ |

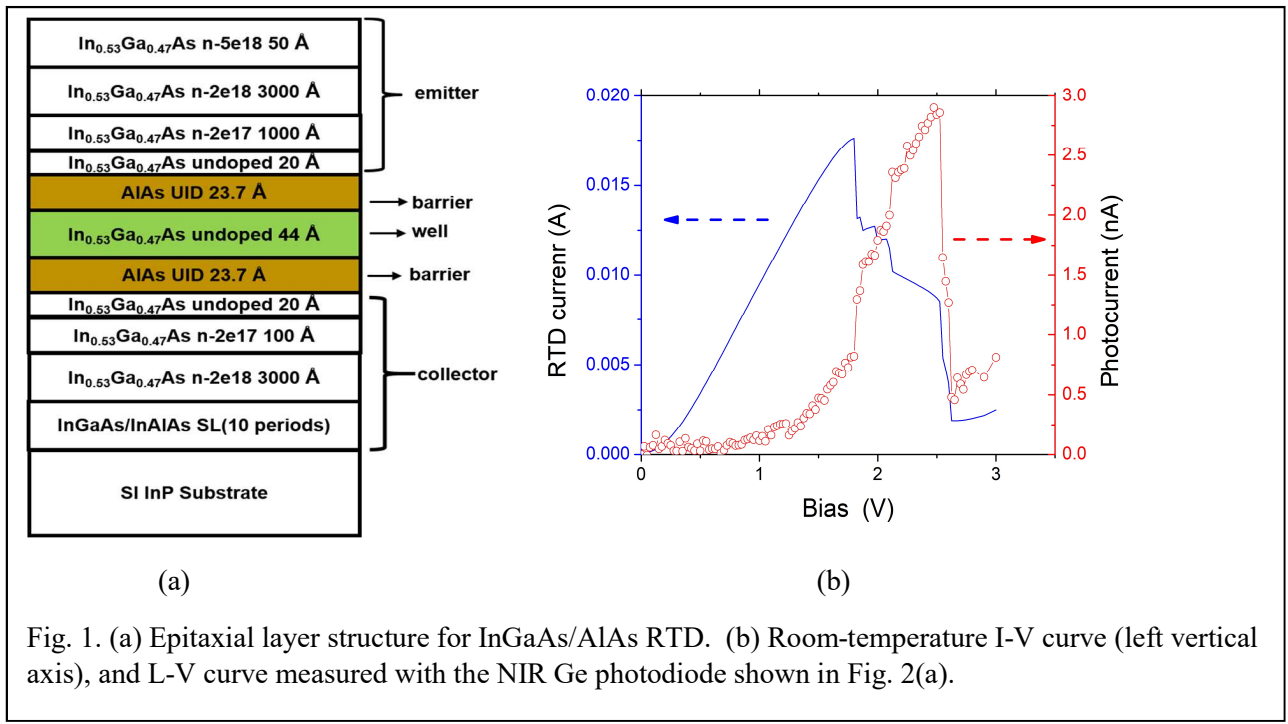

(a)                          (b)

Fig. 1. (a) Epitaxial layer structure for InGaAs/AlAs RTD. (b) Room-temperature I-V curve (left vertical axis), and L-V curve measured with the NIR Ge photodiode shown in Fig. 2(a).

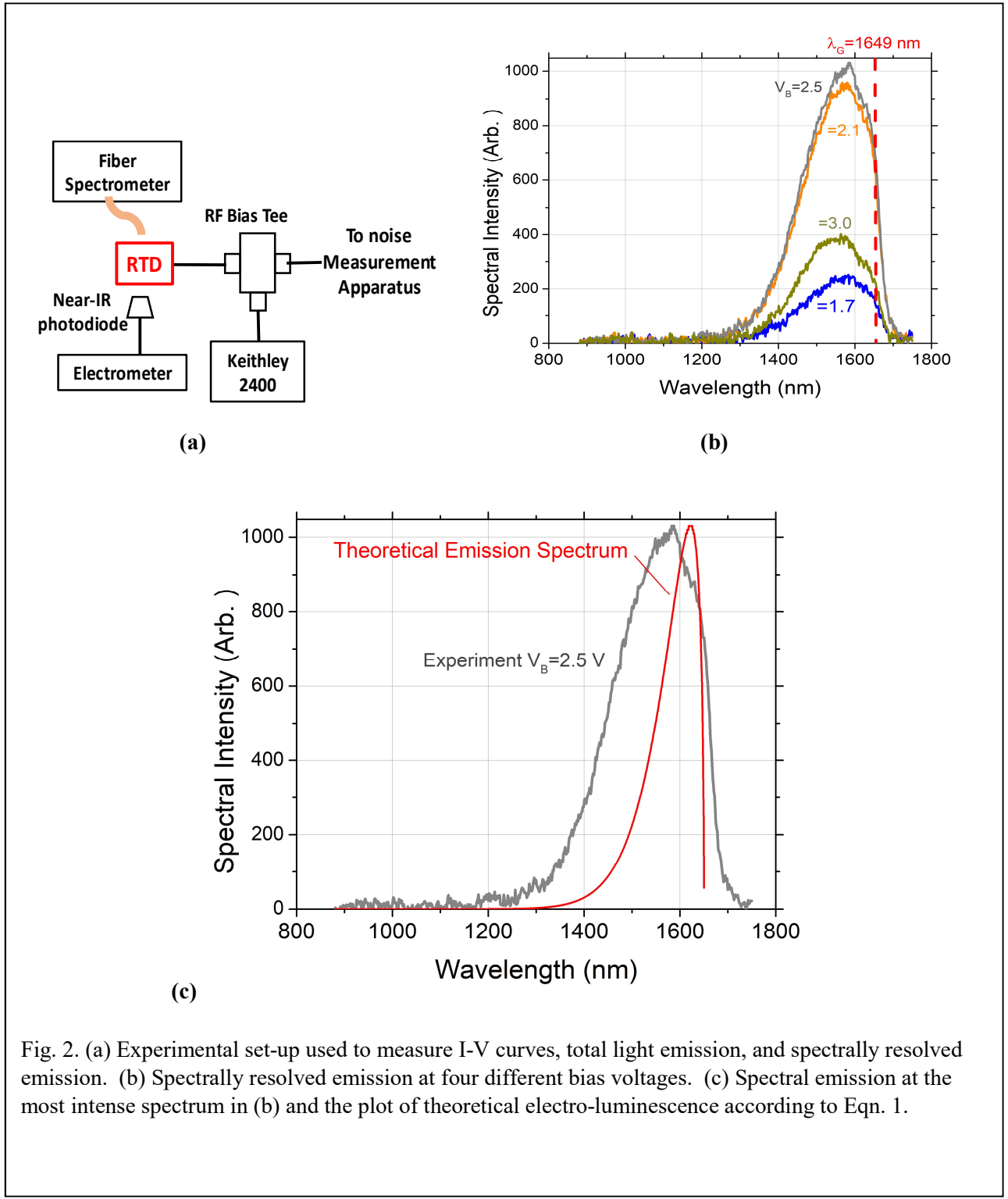

Fig. 2. (a) Experimental set-up used to measure I-V curves, total light emission, and spectrally resolved emission. (b) Spectrally resolved emission at four different bias voltages. (c) Spectral emission at the most intense spectrum in (b) and the plot of theoretical electro-luminescence according to Eqn. 1.

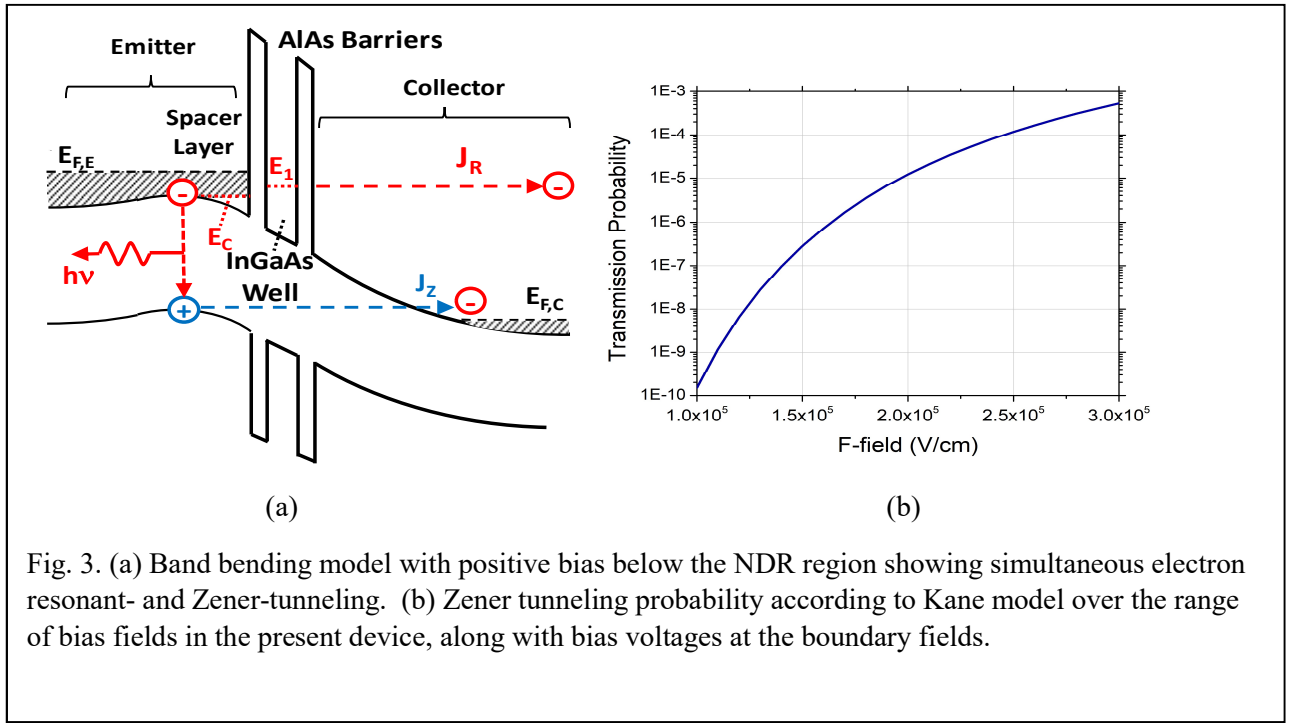

Fig. 3. (a) Band bending model with positive bias below the NDR region showing simultaneous electron resonant- and Zener-tunneling. (b) Zener tunneling probability according to Kane model over the range of bias fields in the present device, along with bias voltages at the boundary fields.

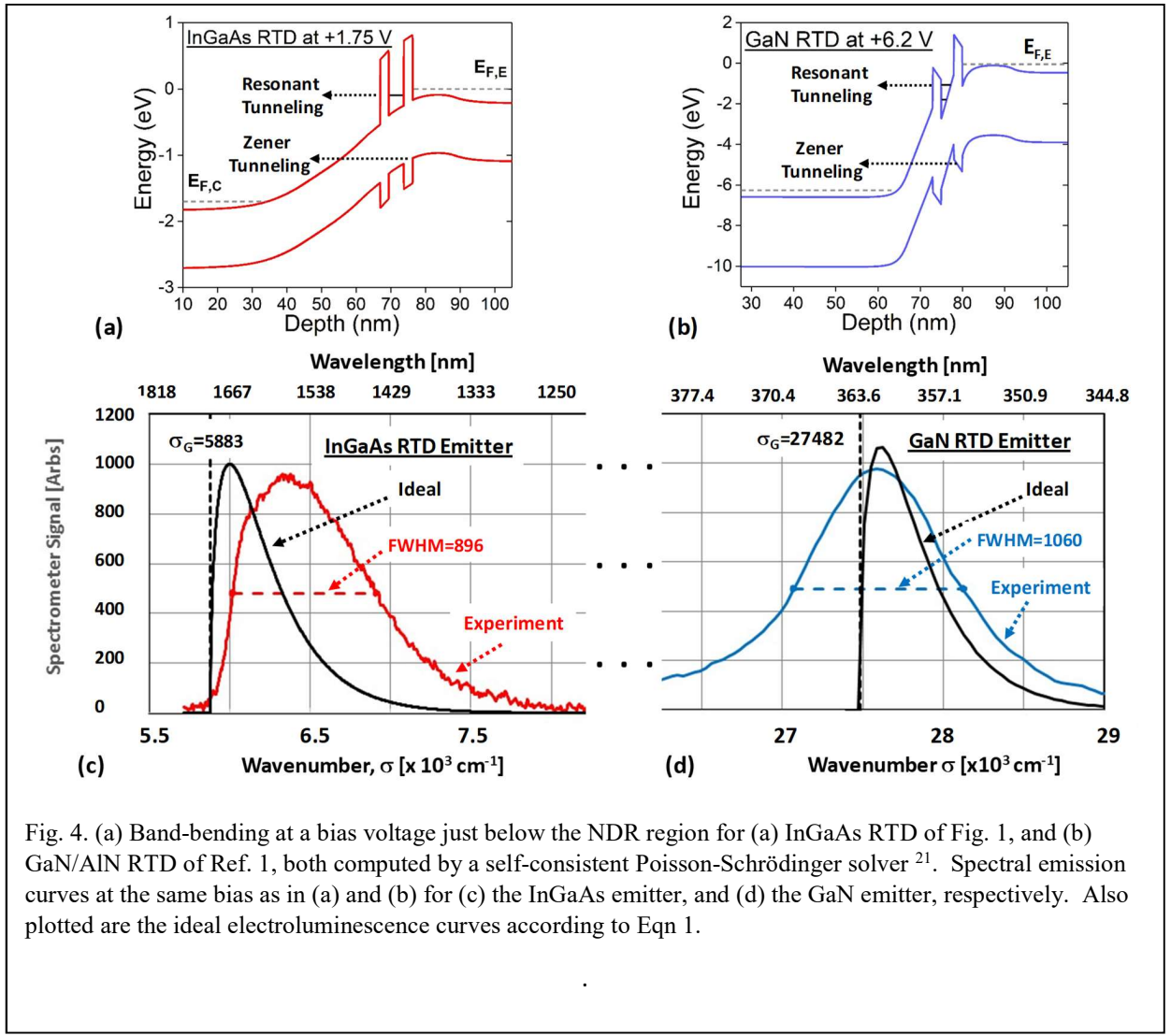

Fig. 4. (a) Band-bending at a bias voltage just below the NDR region for (a) InGaAs RTD of Fig. 1, and (b) GaN/AlN RTD of Ref. 1, both computed by a self-consistent Poisson-Schrödinger solver [21]. Spectral emission curves at the same bias as in (a) and (b) for (c) the InGaAs emitter, and (d) the GaN emitter, respectively. Also plotted are the ideal electroluminescence curves according to Eqn 1.

.